\documentclass{revtex4}
\usepackage{amsmath}

\begin{document}

\title{Brane in 6D with increasing gravitational trapping potential}
\author{Merab Gogberashvili}
\email{gogber@hotmail.com}
\affiliation{Andronikashvili Institute of Physics, 6 Tamarashvili Str., Tbilisi 380077
Georgia}

\author{Douglas Singleton}
\email{dougs@csufresno.edu}
\affiliation{Physics Dept., CSU Fresno, 2345 East San Ramon Ave.
M/S 37 Fresno, CA 93740-8031, USA}

\date{\today}

\begin{abstract}
A new solution to Einstein equations in (1+5)-spacetime with an embedded 
(1+3) brane is given. This solution localizes the zero modes of all kinds of matter
fields and 4-gravity on the (1+3) brane by an increasing,
transverse gravitational potential. This localization occurs despite 
the fact that the gravitational potential is not a decreasing exponential, and 
asymptotically approaches a finite value rather than zero. 
\end{abstract}


\maketitle

The main question of brane models is how to localize fields on the
brane. To localize multi-dimensional fields on the brane the effective
``coupling'' constants appearing after integration of the
Lagrangian over the extra coordinates must be non-vanishing and finite. For reasons 
of economy one would like to have a single, universal trapping mechanism that
works for all fields. It is natural to try a gravitational trapping of the 
physical fields on the brane, since gravity is known to have a universal
coupling with all matter fields.

In (1+4)-dimensional models the following results were established: spin
$0$  and spin $2$ fields are localized on the brane with a decreasing, exponential
gravitational warp factor, spin $1/2$ field are localized on the brane with
an increasing warp factor \cite{RaSu,BaGa}, and spin $1$ fields are not
localized at all \cite{Po}. For the case of (1+5)-dimensions it
was found that spin $0$, $1$ and $2$ fields are localized on the
brane with a decreasing warp factor and spin $1/2$ fields are localized on the
brane with increasing warp factor \cite{Od}. So in both (1+4)-, or
(1+5)-space models one is required to introduce some non-gravitational interaction in order 
to localize the Standard Model particles. 

Here we want to show that zero modes of spin $0$, $1/2$, $1$ and
$2$ fields all can be localized on the brane in a (1+5)-space by an 
increasing warp factor which is not an exponential. A similar solution for a (2+4)-signature 
metric was found in previous work \cite{GoMi}. Having a growing
gravitational potential (warp factor) is opposite to the choice
of the Randall-Sundrum model where the warp factor 
maximum is on the brane \cite{RaSu}. However, Newton's law still holds on the brane 
as a result of the cancellation mechanism introduced in \cite{Go} which
allows both increasing and decreasing types of gravitational potential.

The action of the gravitating system in six dimensions can be written
in the form
\begin{equation} \label{action}
S = \int d^6x\sqrt{- ^6g}\left[\frac{M^4}{2}(^6R + 2 \Lambda) + L
\right] ~,
\end{equation}
where $\sqrt{-^6g}$ is the determinant, $M$ is the fundamental scale,
$^6R$ is the scalar curvature, $\Lambda$ is the cosmological
constant and $L$ is the Lagrangian of matter fields. All of these
quantities are six dimensional. 

Einstein's 6-dimensional equations with stress-energy tensor $T_{AB}$ are
\begin{equation} \label{Einstein6}
    R_{AB} - \frac{1}{2} g_{AB} ~^6R = \frac{1}{M^4}\left(\Lambda
   g_{AB} + T_{AB}\right)~.
\end{equation}
Capital Latin indices run over $A, B,... = 0, 1, 2, 3, 5, 6 $.

We will look for solutions of \eqref{Einstein6} that contain
4-dimensional Minkowski geometry with the following ansatz
\begin{equation} \label{ansatz}
ds^2= \phi ^2(x^i) \eta_{\alpha \beta }(x^\nu)dx^\alpha dx^\beta +
g_{ij}(x^i)dx^i dx^j ~.
\end{equation}
The metric of ordinary 4-dimensional space, $\eta_{\alpha
\beta }(x^\nu)$, has the signature $(+,-,-,-)$. The Greek indices $\alpha,
\beta,... = 0, 1, 2, 3$ refer to the coordinates of these 4-dimensions, 
while small Latin indices $i, j, ... = 5, 6$ refer to coordinates of the
transverse space. It is assumed that ansatz \eqref{ansatz}
only depends on the extra coordinates, $x^i$, through the 4-dimensional conformal
factor, $\phi^2$, and the metric tensor of transversal 2-space,
$g_{ij}$.

Since gravity in the transverse 2-space is trivial, the metric of the extra space,
$g_{ij}(x^i)$, has only one independent component. We will choose
its diagonal component as independent
\begin{equation} \label{gij}
g_{ij}(x^i) = -\delta_{ij} \lambda (r) ~,
\end{equation}
$\delta_{ij}$ is the metric of Euclidean 2-space.

We require cylindrical symmetry of the transverse 2-space. It is 
convenient to write the two extra spatial dimensions $x^5$ and $x^6$ 
using polar coordinates $(r,\theta)$, where $0\leq r
=\sqrt{x^2_5 + x^2_6} < \infty$ and $0\leq \theta < 2\pi$. Thus
ansatz \eqref{ansatz} takes the form
\begin{equation} 
\label{metric}
ds^2 = \phi^2(r) \eta_{\alpha \beta }(x^\nu)dx^\alpha dx^\beta -
\lambda (r)(dr^2 + r^2d\theta^2)~ .
\end{equation}

This metric is slightly different from the metrics investigated in other
(1+5)-space brane models \cite{Od,Gr,GhSh}. The
independent metric function of the extra space, $\lambda (r)$,  serves as a
conformal factor for the Euclidean 2-dimensional metric of the transverse
space just as the function $\phi^2(r)$ does for the 4-dimensional part. Usually in
6-dimensional ans{\"a}tze the independent metric function multiples only the angular part
($d\theta^2$) of the metric \eqref{metric} and corresponds to a cone-like geometry of a
string-like defect with a singularity at the center $r = 0$. 
We want to look for nonsingular solutions, and so we choose the metric to take the 
form \eqref{metric} where the extra part is conformally equivalent to Euclidean 2-space.

The nonzero components of the stress-energy tensor $T _{AB}$ are
assumed to  be
\begin{equation} \label{source}
T_{\mu\nu} = - g_{\mu\nu} F(r), ~~~ T_{ij} = - g_{ij}K(r),
     ~~~ T_{i\mu} = 0 ~.
\end{equation}
The two source functions, $F(r)$ and $K(r)$, are assumed to depend only on
the radial coordinate $r$.

For the 4-dimensional Einstein equations outside the brane we require
the ordinary form without a cosmological term
\begin{equation} \label{Einstein4}
R_{\mu\nu} - \frac{1}{2} \eta_{\mu\nu} R = 0~.
\end{equation}
The Ricci tensor in four dimensions $R_{\alpha\beta}$ is
constructed from the 4-dimensional metric tensor
$\eta_{\alpha\beta}(x^{\nu})$ in the standard way. Then with the ans{\"a}tze 
\eqref{metric} and \eqref{source} the Einstein field
equations \eqref{Einstein6} become
\begin{eqnarray}
\label{Einstein6a}
3 \frac{\phi ^{\prime \prime}}{\phi} + 3 \frac{\phi ^{\prime}}{r \phi} 
+ 3 \frac{(\phi ^{\prime})^2}{\phi ^2} +\frac{1}{2}\frac{\lambda^{\prime \prime}}{\lambda }
-\frac{1}{2}\frac{(\lambda ^{\prime})^2}{\lambda ^2}+\frac{1}{2}\frac{\lambda ^{\prime}}{r\lambda } 
= \frac{\lambda }{M^4} (F(r) - \Lambda ) ~, \nonumber \\
\frac{\phi ^{\prime} \lambda ^{\prime}}{\phi \lambda }
+ 2 \frac{\phi ^{\prime}}{r \phi} +3 \frac{(\phi ^{\prime})^2}{\phi ^2} = 
\frac{\lambda }{2 M^4} ( K(r)-  \Lambda ) ~, \\
2 \frac{\phi ^{\prime \prime}}{\phi}  - \frac{\phi ^{\prime} \lambda ^{\prime}}{\phi \lambda }
+3 \frac{(\phi ^{\prime})^2}{\phi ^2} = 
\frac{\lambda}{2 M^4} (K(r) - \Lambda ) ~, \nonumber
\end{eqnarray}
where the prime $=\partial / \partial r$. These equations are for the $\alpha \alpha$, 
$rr$, and $\theta \theta$ components respectively. 

Subtracting the $rr$ from the $\theta \theta$ equation and multiplying by 
$\phi / \phi ^{\prime}$ we arrive at 
\begin{equation}
\label{phi-g}
\frac{\phi ^{\prime \prime}}{\phi ^{\prime}} - \frac{\lambda  ^{\prime}}{\lambda } -\frac{1}{r} = 0 ~.
\end{equation}
This equation has the solution
\begin{equation}
\label{g}
\lambda  (r)= \frac{\rho^2 \phi ^{\prime}}{r} ~,
\end{equation}
where $\rho$ is an integration constant with units of length. 

We want to find a solution $\phi (r)$ (and thereby
also $\lambda (r)$ via \eqref{g}) which will provide a universal, gravitational trapping for
all kinds of matter fields. In the above the brane is located at the origin $r=0$. 
From \eqref{g} we see that to avoid singularities the function $\lambda \propto \phi ^{\prime}$ 
can not change sign and should approach zero only on the brane $r \rightarrow 0$
and at infinity $r \rightarrow \infty$. Therefore we must look for
a solution $\phi$ which is a monotone function of $r$ and increases
or decreases from the brane to some finite value at infinity. 

A field will be trapped on the brane if  the
integral over the extra coordinates $r, \theta$ converges.
From the metric ansatz \eqref{metric} and the solution for $\lambda (r)$ in \eqref{g}
the general expression for these integrals is 
\begin{equation}\label{planck}
\int_0^{2\pi} d\theta\int_0^\infty dr~ \phi ^c (\phi ^{\prime})^d \sqrt{-^6g} =
2\pi \int_0^\infty dr~ \phi ^c (\phi ^{\prime})^d r\phi^4 \lambda  \sqrt{-\eta}  = 2 \pi\rho^2 \int_0^{\infty}
dr \phi^n (\phi ^{\prime})^m  \sqrt{-\eta}~,
\end{equation}
where $\eta$ is the determinant of ordinary 4-dimensional metric tensor,
$c$ and $d$ are numbers that depend on the type of field being trapped, and
$n=c+4$ and $m=d+1$.  To have localized physical
fields on the brane the integral of $\phi^n (\phi ^{\prime})^m$ must go to zero fast enough as
$r \rightarrow \infty$, and not have any essentially singularities in the range
$0 \le r \le \infty$. For the fields that we will consider (spin-0, spin-1/2, spin-1, and spin 2)
$m$ is positive while $n$ can be either positive or negative. 
One standard possibility is to have an exponential warp factor, 
$\phi (r) \propto e^{-cr}$ \cite{Od, Gr, GhSh}. In this case for fields with $n>0$ 
$\phi ^n$ and $(\phi ')^m$ go to zero as $r \rightarrow \infty$ and the integral
\eqref{planck} converges and the fields will be trapped. However for $n<0$ (for 
example with spinor fields) the integral diverges and the fields are not trapped. 

In this work we present a different type of solution where the 
gravitational potential, $\phi$, is not an exponential and goes to a 
finite value at $r=\infty$ ({\it i.e.} $\phi(\infty) = const$), yet nevertheless is able
to trap the zero modes of all types of fields since $\phi ' \rightarrow 0$. 
A  similar non-exponential solution for the signature (2+4) was
found in previous work \cite{GoMi}. This type of solution
is impossible in the 5-dimensional case where only an
exponential warp factor is possible.  

Based on the above discussion we choose the boundary conditions  on the brane as
\begin{equation} \label{r=0}
\phi (r\le \epsilon) \approx 1 ~, ~~~ \phi^\prime (r\le \epsilon)
\approx 0~,
\end{equation}
where $\epsilon$ is the brane width. The boundary conditions at infinity are
\begin{equation} \label{asimptotics}
\phi (r \rightarrow \infty) \rightarrow a  ~ ,~~~~~\phi^\prime (r
\rightarrow \infty) \rightarrow 0 ~,
\end{equation}
where $a>1$ is some constant. Since the function $\phi^\prime$ is proportional to the metric of
the extra 2-space, $g_{ij}$, the boundary conditions \eqref{r=0} and
\eqref{asimptotics} imply that on the brane and at infinity
the effective geometry of space-time is 4-dimensional.

The energy-momentum conservation equation 
\begin{equation}
\label{energy-con}
\nabla ^A T_{AB} = \frac{1}{\sqrt{-^6 g}}
\partial _A (\sqrt{-^6 g}T^{AB}) + \Gamma ^B _{CD} T^{CD} = 0
\end{equation}
gives a relationship between the two source functions $F(r)$ and $K(r)$ 
in \eqref{source}. In terms of the ansatz functions the conservation law reads 
\begin{equation} \label{deltaT}
K^{\prime} + 4\frac{\phi^\prime}{\phi} \left(K - F \right) = 0 ~.
\end{equation}
This equation places a restriction on the form of the source functions, which will
allow us to determine a simple form for $F(r)$ and $K(r)$. We assume that they
are smooth functions of the radial coordinate $r$ and describe a continuous matter distribution
within a core of radius $\epsilon$. We also assume that they decrease rapidly when 
$r>\epsilon $. Below we will find that the transverse gravitational potential, 
$\phi^2(r)$, can be a growing function as one moves off
the brane, so the factor $1/\phi ^2(r)$ has $\delta$-like behavior. 
Thus outside the core we will set the source
functions proportional to $1/\phi^2$. This form satisfies equation \eqref{deltaT} with
the sources functions given by
\begin{equation} \label{FK}
F(r>\epsilon ) = \frac{f}{2 \phi^2}~, ~~~
K(r>\epsilon ) = \frac{f}{\phi^2} ~,
\end{equation}
where $f$ is some constant.

Now we shall find the outer solution ($r > \epsilon$) of the 6-dimensional
Einstein equations \eqref{Einstein6} for the brane when the
metric and matter energy-momentum tensor have the general
forms given in \eqref{metric} and \eqref{FK} respectively. 
The system \eqref{Einstein6a}, after the insertion of \eqref{g} and \eqref{FK}, 
has only one independent equation. Taking either the $rr$ or $\theta \theta$ component
of these equations and multiplying by $r \phi ^4$ gives
\begin{equation}
\label{rr}
r \phi ^3 \phi ^{\prime \prime} +\phi ^3 \phi ^{\prime} + 3 r \phi ^2 (\phi ^{\prime} )^2 =
\frac{\rho ^2 \phi ^4 \phi ^{\prime}}{2 M^4} \left( \frac{f}{\phi ^2} - \Lambda \right) ~.
\end{equation} 
Taking the first integral of this equation and setting the integration constant to zero yields \cite{GoMi}
\begin{equation} \label{first}
r \phi^\prime = A(-\phi^2 + a^2) ~,
\end{equation}
where we have introduced the parameters
\begin{equation} \label{parameters}
A = \frac{\rho^2 \Lambda}{10 M^4} ~ ,~~~~~ a^2 = \frac{5f}{3\Lambda}~.
\end{equation}
From the equation \eqref{first} we see that to have an increasing
function $\phi$ ($\phi ^{\prime} >0$) we must require the following conditions
\begin{equation} \label{constants}
a>\phi ~,~~~~~ \Lambda > 0~ ,~~~~~ f >0  ~.
\end{equation}
Equation \eqref{first} is easy to integrate
\begin{equation} \label{phi}
\phi = a\tanh \left( b \ln \left( \frac{r}{c} \right) \right) = a\frac{r^{2b} - c^{2b}}{r^{2b} + c^{2b}} ~,
\end{equation}
where $b = Aa$. The integration constant $c$ can be fixed from
the boundary conditions \eqref{r=0}
\begin{equation} \label{c}
c^{2b} = \epsilon^{2b} \frac{a - 1}{a + 1} ~,
\end{equation}
where $\epsilon$ is the brane width.

It is easy to show that 4-dimensional gravity is localized on
the brane by the solution \eqref{phi} in spite of its growing character.
We will only consider spin-2 modes, which are transverse,
traceless fluctuations $H_{\mu\nu}$ around the background metric
\eqref{metric}
\begin{equation} \label{metric+}
ds^2 = \phi^2(r) (\eta_{\mu\nu}+ H_{\mu\nu})dx^\mu dx^\nu  -
\frac{\rho^2 \phi^{\prime}}{r}(dr^2 + r^2d\theta^2)~ ,
\end{equation}
where $\phi$ has the form \eqref{phi}.

In the gauge $\nabla^\mu H_{\mu\nu} = H = 0$ Einstein's equations
reduce to the form of the linearized equations
\begin{equation} \label{fluc}
\frac{1}{\sqrt{-g}} \partial_A (\sqrt{-g} g^{AB} \partial_B
H_{\mu\nu}) - 2 \Lambda H_{\mu\nu} = 0~.
\end{equation}
We look for solutions of the form
\begin{equation}
H_{\mu\nu}(x^A) = h_{\mu\nu}(x^\mu) \sum_{lm} \sigma_m(r) e^{i l \theta}~,
\end{equation}
where $h_{\mu\nu}(x^\mu)$ satisfy the 4-dimensional field equations
\begin{equation}
\label{brane-fluc}
(\Delta - 2 \Lambda \phi^2)h_{\mu\nu}(x^\mu) = - m_0^2
h_{\mu\nu}(x^\mu) ~,
\end{equation}
with the definition of 
\begin{equation}
\label{laplace4}
\Delta = \frac{1}{\sqrt{-\eta}}\partial_\mu
(\sqrt{-\eta} \; \; \eta^{\mu\nu}\partial_\nu) .
\end{equation}  

Note that the presence of the cosmological constant term,  
$2 \Lambda \phi^2$ in \eqref{brane-fluc} which appears in most studies of
the localization of spin 2 fields \cite{RaSu, Od, Gr, GhSh}, is problematic since it implies a
negative energy, tachyonic spin-0 mode for the fluctuation, $h_{\mu \nu}$ on the brane.
One could consider fluctuations coming from the extra coordinates, which
might have a chance to cancel this contribution. Without a satisfactory quantum theory
of gravity it is not certain how to the properly  resolve this difficulty, and we
will not consider its resolution here.
 
Using \eqref{brane-fluc} equation \eqref{fluc} reduces to
\begin{equation}
{\sigma ^{\prime \prime}} + \left(\frac{1}{r} +
4~\frac{\phi ^{\prime}}{\phi}\right)~{\sigma ^{\prime}} +
\left(\frac{m^2 _0 \rho^2}{r} ~ \frac{{\phi ^{\prime}}}{\phi^2} - \frac{l^2}{r^2}\right) ~ \sigma = 0 ~,
\end{equation}
It is easy to see that this equation has the zero-mass ($m_0 = 0$) and s-wave
($l = 0$) constant solution $\sigma_0 = const$. Substitution of
this zero mode into the Einstein-Hilbert action leads to
\begin{equation}
S^{(0)} \sim  \sigma_0^2 \int_0 ^{2 \pi} d\theta \int_0^{\infty} dr \frac{\sqrt{-^6g}}{\phi^2}
\int d^4 x [\partial^\alpha h^{\mu\nu}
\partial_\alpha h_{\mu\nu} + \cdots ] ~ .
\end{equation}
To have the 4-dimensional spin-2 graviton localized on the brane requires that the integral over
$r$ and $\theta$ (which corresponds to the 4-dimensional Planck scale) converge
\begin{eqnarray}
m^2 _{Pl} = 2 \pi M^4 \int_0^{\infty} dr \frac{r \phi ^4 \lambda}{\phi^2} = 2 \pi \rho ^2 M^4 \int_0^{\infty}
\phi^2 \phi^\prime dr = 2 \pi \rho ^2 M^4\int_1^{a} \phi^2 d\phi =\frac{2}{3} \pi \rho ^2 M^4 (a^3-1) ~.
\end{eqnarray}

Now we want to check that zero-modes of matter fields also are
localized on the brane with the non-exponential, increasing warp factor \eqref{phi}.

For spin-0 fields in six dimensions we assume that the fields are independent
of the extra coordinates so that the action can be cast in the form
\begin{equation} \label{action0}
S_\Phi = -\frac{1}{2} \int d^6x \sqrt{-^6g}g^{AB}\partial _A \Phi
\partial _B \Phi = -\pi \rho^2 \int_1^a
\phi^2d\phi \int d^4x \sqrt{-\eta}\eta^{\mu\nu}\partial _\mu
\Phi\partial _\nu \Phi ~.
\end{equation}
The integral over the extra coordinates in \eqref{action0} is the same as for the spin-2 case above.
Thus the integral is finite and the spin-0 field is localized on the brane.

The action for a $U(1)$ vector, gauge field in the case of
constant extra components ($A_i = const $) reduces to the
4-dimensional Maxwell action multiplied an integral over the
extra coordinates
\begin{equation} \label{action1}
S_A = -\frac{1}{4} \int d^6x \sqrt{-^6g}g^{AB} g^{MN}F_{AM}F_{BN} =
-\frac{\pi}{2} \rho^2 \int_1^a d\phi \int d^4x
\sqrt{-\eta}\eta^{\mu\nu}\eta^{\alpha\beta}F_{\mu\alpha}F_{\nu\beta}~.
\end{equation}
This integral is also finite, and the gauge field is localized on the brane.

In the case of spinor fields we introduce the sechsbein
$h^{\bar{M}}_M$, where $\bar{M}, \bar{N}, ...$ denotes local
Lorentz indices. The spin connection is defined as
\begin{equation} \label{spin1}
\omega^{\bar{M}\bar{N}}_M = \frac{1}{2} h^{N\bar{M}} (\partial_M
h^{\bar{N}}_N - \partial_N h^{\bar{N}}_M) - \frac{1}{2}
h^{N\bar{N}}(\partial_M h^{\bar{M}}_N - \partial_N h^{\bar{M}}_M)
- \frac{1}{2} h^{P\bar{M}}h^{Q\bar{N}}(\partial_P h_{Q\bar{R}} -
\partial_Q h_{P\bar{R}})h^{\bar{R}}_M ~.
\end{equation}
The non-vanishing components of the spin-connection for the
background metric \eqref{ansatz} are
\begin{equation} \label{spin2}
\omega^{\bar{r}\bar{\nu}}_\mu = \delta ^{\bar{\nu}} _{\mu} \frac{\sqrt{r \phi '}}{\rho} ~ , ~~~~~
\omega^{\bar{r}\bar{\theta}}_\theta = \sqrt{\frac{r}{\phi '}} \partial _r 
\left( \sqrt{r \phi' } \right) ~.
\end{equation}

Therefore, the covariant derivatives have the form
\begin{equation} \label{covariant}
D_\mu\Psi = \left( \partial_\mu +  \frac{1}{2} \omega^{\bar{r}\bar{\nu}}_\mu 
\gamma _r \gamma _\nu \right) \Psi ~ , ~~~~~ D_r\Psi = \partial_r \Psi ~ , ~~~~~
D_\theta \Psi = \left(\partial_\theta + \frac{1}{2} \omega^{\bar{r}\bar{\theta}}_\theta 
\gamma _r \gamma _\theta \right) \Psi ~.
\end{equation}
where $\gamma _{\nu} , \gamma _r , \gamma _{\theta}$ are gamma matrices.

We are looking for solutions of the form $\Psi (x^A) = \psi
(x^\nu)B(r)$, where $\psi $ satisfies the massless 4-dimensional
Dirac equation, $\gamma^\nu \partial_\nu \psi = 0$ . Then the
6-dimensional massless Dirac equation
\begin{equation}
\Gamma ^{\mu} D_{\mu} \Psi + \Gamma ^r D_r \Psi +
\Gamma ^{\theta} D_{\theta} \Psi =0 ~,
\end{equation}
where $\Gamma ^A = h^A _{\bar {B}} \gamma ^B$ are the
6-dimensional curved space gamma matrices, reduces to
\begin{equation} \label{equationB}
\left[ \partial_r + 2 \frac{\phi '}{ \phi} + \frac{1}{2} \frac{1}{\sqrt{r \phi'}} \partial_r 
\left( \sqrt{r \phi ' } \right)  \right] B(r) = 0 ~.
\end{equation}
The solution of this equation with the integration constant, taken as one, is
\begin{equation} \label{B}
B(r) = \phi^{-2} \left( r \phi ' \right)^{-1/4} .
\end{equation}
Using the solution from \eqref{phi} one finds that $B(r) \rightarrow \infty$ as $r \rightarrow
\infty$. This happens since $r \phi ' \rightarrow 0$ as $r \rightarrow \infty$. This appears
to imply that the total spinor wavefunction, $\Psi (x^A)$, does not have good asymptotic behavior.
However, the effective wavefunction is $\Psi (x^A) (- ^6 g ) ^{1/4} \propto B(r) \sqrt{\rho ^2 \phi ^4 \phi '}
\rightarrow \rho (\phi ' / r)^{1/4}$, which goes to zero as $r \rightarrow \infty$. 
The action of the spin $1/2$ field takes the form
\begin{equation} \label{action1/2}
S_\Psi = \int d^6 x \sqrt{-^6g}\bar{\Psi} i \Gamma^A D_A \Psi =
2\pi \rho^2 \int_0^\infty dr~ r^{-1/2} \phi^{-1} (\phi ')^{1/2}\int d^4 x
\sqrt{-\eta } \bar{\psi} i \gamma^\nu
\partial_\nu \psi ~ .
\end{equation}
Using the explicit form of the solution \eqref{phi} one finds that
the integral over $r$ in \eqref{action1/2}
is of the form
\begin{equation}
\label{fermi-int}
\int dr~ \sqrt{\frac{\phi '}{r \phi^2}} = \frac{1}{\sqrt{ab}} \ln \left(\frac{r^b - c^b}{r^b+c^b} \right)
\end{equation}
This integral converges if one evaluates it from $r=\epsilon$ to $r=\infty$. 
Thus the massless Dirac fermions are also localized on the brane.

When we consider interaction of scalars, or fermions with the
electromagnetic field we must make the usual replacements
\begin{equation} \label{interaction}
\partial_i \rightarrow \partial_i - iA_i , ~~~~~\Psi \rightarrow
e^{iA_i x^i} \Psi
\end{equation}
in above formula for localization. Here $x^i$ are coordinates of
transverse 2-space and $A_i$ are constant extra components of
electromagnetic field.

To summarize, in this paper it is shown that for a realistic
form of the brane stress-energy, there exists a non-singular
static solution of 6-dimensional Einstein equations. This solution
provides gravitational trapping of the 4-dimensional gravity and
matter fields on the brane without extra $\delta$-like sources. In
contrast to the Randall-Sundrum case, the factor responsible for
this trapping is an increasing gravitational potential.
Despite this and the fact that the transverse space is infinite, the 
integral over the extra coordinates is convergent, and the fields are localized on the
brane. 

\begin{flushleft}
{\bf Acknowledgments} This work is supported by a 2003 COBASE grant.
\end{flushleft}

\end{document}